# High-Performance Parallel Optimization of the Fish School Behaviour on the Setonix Platform Using OpenMP

Haitian Wang (23815631) Long Qin (23829101)

## Abstract

This paper presents an in-depth investigation into the high-performance parallel optimization of the Fish School Behaviour (FSB) algorithm on the Setonix supercomputing platform using the OpenMP framework. Given the increasing demand for enhanced computational capabilities for complex, large-scale calculations across diverse domains, there's an imperative need for optimized parallel algorithms and computing structures. The FSB algorithm, inspired by nature's social behavior patterns, provides an ideal platform for parallelization due to its iterative and computationally intensive nature. This study leverages the capabilities of the Setonix platform and the OpenMP framework to analyze various aspects of multi-threading, such as thread counts, scheduling strategies, and OpenMP constructs, aiming to discern patterns and strategies that can elevate program performance. Experiments were designed to rigorously test different configurations, and our results not only offer insights for parallel optimization of FSB on Setonix but also provide valuable references for other parallel computational research using OpenMP. Looking forward, other factors, such as cache behavior and thread scheduling strategies at micro and macro levels, hold potential for further exploration and optimization.

## 1 Introduction

### 1.1 Background

With the rapid advancements in scientific research and technology, there has been an escalating demand for computational power, especially for large-scale and intricate calculations. This is evident across various domains, including biology, physics, meteorology, and engineering[1] . In this context, high-performance computing (HPC) plays an instrumental role, not just by meeting these computational needs but, more critically, by enhancing computational speeds through optimized parallel algorithms and frameworks[2] . For instance, the acceleration in weather forecasting models ensures quicker predictions for the public, while intricate simulations in drug development yield outcomes in reduced time frames[3] .

Particle Swarm Optimization (PSO) is an adaptive optimization technique inspired by the social behavior patterns observed in nature. A variant of this is the Fish School Behaviour (FSB) algorithm, which emulates the behavior of fish as they search for food, guiding the majority to areas abundant in food. Given its iterative, repetitive, and computationally intensive nature, FSB stands out as an ideal candidate for parallelization, offering a robust experimental platform for HPC research[4] .

OpenMP, a multi-platform API, has emerged as a popular parallel programming approach, especially well-suited for modern multi-core and multi-threaded processor architectures[5] . Its simplicity and flexibility make it an attractive choice for intensive computations[6] .

The Setonix supercomputing platform, affiliated with the University of Western Australia (UWA), serves as a state-of-the-art computing resource for the academic community. Its compatibility with OpenMP allows researchers to effortlessly deploy and test parallel algorithms and applications based on this framework.

## 1.2 Objectives

The central objective of this study is to harness the computational prowess of the Setonix platform and the OpenMP framework to conduct high-performance parallel optimization of the FSB algorithm, with a primary focus on intricate, recurring, and large-scale computational tasks.

Our analysis dives deep into the impact of various multi-threading facets on program performance on the Setonix platform, zeroing in on execution times. We will examine the implications of different thread counts, scheduling strategies, and OpenMP constructs on efficiency. By undertaking a series of qualitative and quantitative experiments, we aim to discern underlying patterns and identify strategies and techniques that can notably elevate program performance.

Through this research endeavor, we aspire to shed light on the parallel optimization of the FSB algorithm on the Setonix platform. Furthermore, we hope our findings serve as valuable references for other parallel computational research on Setonix using OpenMP, contributing to practical solutions for real-world challenges.

## 2 Methodology

This study delves into the exploration of optimization strategies in parallel computing by executing a series of experiments. Initial configurations, including the initial state of fish populations, the simulation of ponds, and the system parameter settings, are vital to commence the investigations. Subsequently, an in-depth analysis of the OpenMP parallel computing framework is conducted, formulating parallelization strategies based on the definition and selection of independent variables. Each segment is delineated in detail below.

## 2.1 Experimental System Configuration

Ensuring consistent starting conditions across experiments is pivotal to warranting the comparability of the results.

### 2.1.1 Initialization and Characteristics of Fish Population

Fish Structure: In the simulation, each fish is characterized by a 'Fish' structure encompassing:

x, y: Denotes the fish's position in a two-dimensional pond.

weight: Represents the fish's weight, affecting its mobility.

prevObjective & currentObjective: They chronicle the fish's objectives at consecutive moments, facilitating behavioral evaluation.

Initialization: At the onset of the simulation, both the initial position and weight of the fish populations are ascertained via random number generators, ensuring simulation diversity. The fish's initial currentObjective is determined by its Euclidean distance from the pond center, providing a starting evaluation metric.

### 2.1.2 Pond Simulation Environment Configuration

Two-dimensional space: The pond simulation manifests as a 200x200 plane, defined by the pond_size parameter.

Fish Behavior: Each fish possesses the freedom to navigate this space. Their directional movement and distance are randomized, yet influenced by their prevailing weight and objectives.

### 2.1.3 System Parameter Configuration

Key experimental parameters, like fish population, initial weight, pond dimensions, and total iterations, are encompassed within the Config structure. Additional parameters, including num_threads, schedule, construct, and chunk_size, which are integral to the parallel strategy, are expounded upon in subsequent sections.

## 2.2 Parallelization Strategy

To enhance simulation efficiency, the OpenMP framework, known for its proficiency in multi-threaded parallelization, is employed. This strategy stems primarily from two considerations: its efficiency in distributing tasks across multiple threads to maximize multi-core processor capabilities, and its ease of transforming serial code into parallel code by simply appending a few compiler directives.

### 2.2.1 Introduction to the OpenMP Framework

OpenMP (Open Multi-Processing) epitomizes a parallel programming model tailored for applications executed on multi-core, multi-processor shared memory computers[6] . Its cardinal features encapsulate:

Ease of Use: Through simple #pragma compiler directives, developers can dictate which code sections should be executed in parallel, seamlessly transitioning from serial to parallel coding.

Flexibility: OpenMP offers an array of APIs, including thread count settings, defining private/shared variables, and specifying synchronization strategies for parallel regions.

Dynamic Thread Management: This allows dynamic thread count adjustments in real-time, adapting to system conditions.

Through OpenMP, loops are parallelized, data is synchronized, and dynamic scheduling policies are executed, with particular emphasis on maintaining data accuracy and consistency during fish movement and feeding operations[7] .

### 2.2.2 Independent Variable Definition & Selection

In-depth exploration of parallelization strategies necessitates scrutiny of specific independent variables:

Thread Count (num_threads): Determine concurrency level in parallel computing.

Scheduling Strategy (schedule): Directs the allocation of parallel tasks, such as static, dynamic, guided, and runtime.

Block Size (chunk_size): In dynamic and guided scheduling, it dictates the volume of tasks assigned to a thread at any given time.

Parallel Structure (construct): Specifies which OpenMP structure is deployed for data synchronization, like reduction or critical.

These variables not only underpin the intricate details of parallel computing but also significantly affect simulation efficiency.

## 2.3 Experimental Design

Utilizing the OpenMP parallel computing paradigm, critical independent variables are scrutinized to discern their impact on holistic system performance, especially the system runtime.

### 2.3.1 Benchmark Settings

To ensure the experiments' validity and comparability, an initial benchmark setting is established:

- $x\_1$: 8
- $x\_2$: schedule(Static, 50)
- $x\_3$: Parallel Construct
- $x\_4$: default

This benchmark acts as a touchstone, offering clarity on how variations in independent variables can influence system performance.

### 2.3.2 Experimental Objectives and Design

Leveraging the capabilities of the Setonix supercomputing platform, our objective is to conduct an in-depth qualitative and quantitative analysis of the four independent variables. The goal is to identify potential optimal combinations to maximize performance, specifically by minimizing the program's runtime. For this purpose, we designed the following three experiments:

**Experiment 1: Impact of Thread Count on Program Runtime**

In this experiment, we predominantly explored the relationship between thread count and program runtime. We considered 16 distinct thread counts, ranging from 1 to 2^16. Based on this, we established four different combinations of scheduling strategies and parallel structures:

- "x_2: schedule(static, 50), x_3: Reduction Construct"
- "x_2: schedule(dynamic, 50), x_3: Reduction Construct"
- "x_2: schedule(guide, 50), x_3: Reduction Construct"
- "x_2: schedule(static, 50), x_3: Critical Construct"

**Experiment 2: Impact of Scheduling Strategy on Program Runtime**

The objective of this experiment is to understand how various scheduling strategies affect the program's runtime. We examined four scheduling strategies: static, dynamic, guided, and runtime. For the static, dynamic, and guided strategies, we considered all potential chunk sizes from 1 to 2^12. For the runtime scheduling strategy, the chunk value wasn't applicable. For each strategy, we recorded the program runtime at three thread count configurations: 2, 4, and 8.

**Experiment 3: Impact of Parallel Constructs on Program Runtime**

In this experiment, our focus is on discerning the differences between the Reduction Construct and the Critical Partial Construct, especially under varying combinations of thread count and scheduling strategy. To elucidate the efficiency of these constructs, we also provided code snippets from the optimized program. In the Reduction Construct, we utilized the "reduction" parallel directive, while the Critical Partial Construct employed the "critical" directive. Here are the code segments for both the reduction and critical implementations:

**Reduction Code:**

```
void eat() {
        double startTime = omp_get_wtime();
        double maxDiff = -INFINITY;
        #pragma omp parallel for reduction(max: maxDiff)

        for(int i = 0; i < NUM_FISH; i++) {
                double diff = school[i].prevObjective -
                school[i].currentObjective;
                if(diff > maxDiff) {
                        maxDiff = diff;
                }
        }
        double endTime = omp_get_wtime();
        time_reduction = endTime - startTime;
}
```

**Critical Code:**

```
void eat() {
        double startTime = omp_get_wtime();
        double maxDiff = -INFINITY;
        #pragma omp parallel for
        for(int i = 0; i < NUM_FISH; i++) {
                double diff = school[i].prevObjective -
                school[i].currentObjective;
                #pragma omp critical
                if(diff > maxDiff) {
                        maxDiff = diff;
                }
        }
        double endTime = omp_get_wtime();
        time_reduction_eat = endTime - startTime;
}
```

# 3 Experimental Results and Data Analysis

## 3.1 Experimental Configuration

The configuration for our experimental environment is shown in Table 1.

**Table 1: Experimental Environment Configuration**

| Configuration Item | Information/Version |
|---|---|
| Operating System Info | SUSE Linux Enterprise Server 15 SP4 15.4 |
| Kernel Version | Linux setonix-01 5.14.21-150400.24.46_12.0.73-cray_shasta_c #1 SMP Thu May 18 23:03:34 UTC 2023 (9c4698c) x86_64 |
| CPU Information | AMD EPYC 7713 64-Core Processor |

| Compiler Version (e.g., GCC) | Copyright (C) 2022 Free Software Foundation, Inc. |
|---|---|
| **Installed Libraries** | **gcc:** <br> The system GNU C Compiler <br> **gcc-c++:** <br> The system GNU C++ Compiler <br> **gcc-fortran:** <br> The system GNU Fortran Compiler <br> **libgomp1:** <br> The GNU compiler collection OpenMP runtime library |

## 3.2 Experiment 1: Influence of Thread Count on Program Runtime

This experiment investigated the variation in program runtime under different thread counts, particularly with constant scheduling strategies and parallel structures. We explored configurations from 1 to 16384 threads. Charts derived from experimental data provided a visual representation of the trends in program runtime across different thread counts. Here's our analysis and conclusion:

### 3.2.1 Reduction Construct vs. Critical Construct with Static Scheduling

First, our attention was directed towards the influence of Reduction Construct and Critical Construct on runtime under a static scheduling strategy. The data indicates that, with a gradual increase in thread count, programs employing Reduction Construct typically had shorter runtimes than those using Critical Construct. This suggests that the Reduction Construct is more efficient than the Critical Construct under the stipulated scheduling strategy. Specifically, the efficiency advantage of the Reduction Construct became more pronounced at medium thread counts like 64 or 128. Although this gap reduced with even higher thread counts (e.g., 8192 and 16384), the Reduction Construct still maintained a shorter execution time. Further detailed analysis of the Critical Construct from Figure 3 revealed that the runtime remains constant with increasing thread counts. Yet, using the Critical Construct, the runtime exhibited a linear growth with increased threads up to 100, then linearly decreased past 100 threads. Upon reaching 800 threads, a minimal value was attained, followed by a slow linear increase, as depicted in Figure 1.

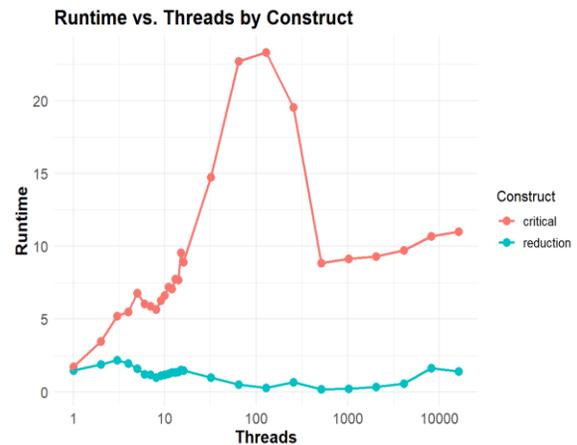

**Fig 1. Runtime vs. Threads by Construct**

### 3.2.2 Static vs. Dynamic vs. Guided with Reduction Structure

We further scrutinized the runtimes under three distinct scheduling strategies—static, dynamic, and guided—within the context of the Reduction Construct. The data displayed a pronounced efficiency of dynamic or guided strategies over the static one, improving performance by up to 50%, especially when the thread count exceeded 20. Specifically, for thread counts under 20, the differences in runtimes across the strategies were marginal, occasionally alternating in momentary dominance. However, beyond 20 threads, dynamic and guided strategies consistently outperformed the static one. Further examination revealed that for counts under 20 threads, runtimes fluctuated around 12 seconds regardless of the strategy. But starting from 20 threads, the runtime began to drop significantly and gradually stabilized. This behavior was observed in both Reduction and Critical Constructs, albeit with different stabilization times, as shown in Figures 2 and 3.

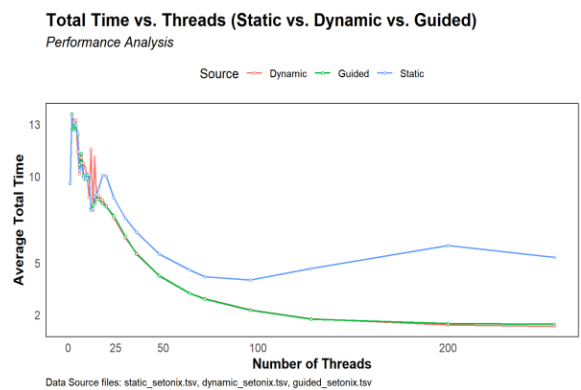

**Fig 2. Runtime vs. Threads (Static vs. Dynamic vs. Guided)**

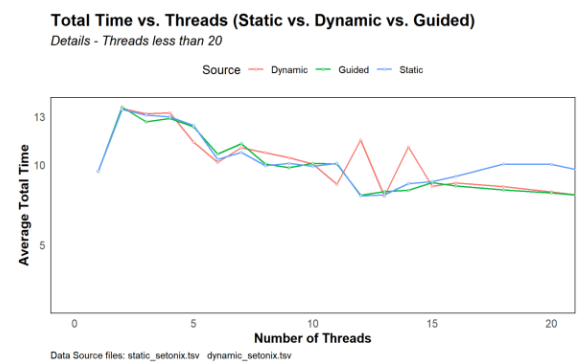

**Fig 3. Runtime vs. Threads (Static vs. Dynamic vs. Guided) - threada less than 20**

## 3.3 Experiment 2: Influence of Scheduling Strategy on Program Runtime

In this section, we extensively examined the impact of varying scheduling strategies and chunk values on program runtime, keeping thread count and parallel constructs constant. Each combination of scheduling strategies recorded the program's execution time. Our analysis and conclusion on the performance impact of different scheduling strategies are as follows:

### 3.3.1 Analysis of Scheduling Strategy's Impact on Runtime

Initially, we considered how different scheduling strategies influenced the execution time of the Reduction Construct at various thread counts. As illustrated in Figure 4, while the runtime of programs using dynamic scheduling remained nearly consistent regardless of chunk size increase, those with static and guided strategies exhibited fluctuations within certain thread segments. Moreover, given identical conditions, dynamic, guided, and runtime strategies performed similarly and were superior to the static strategy.

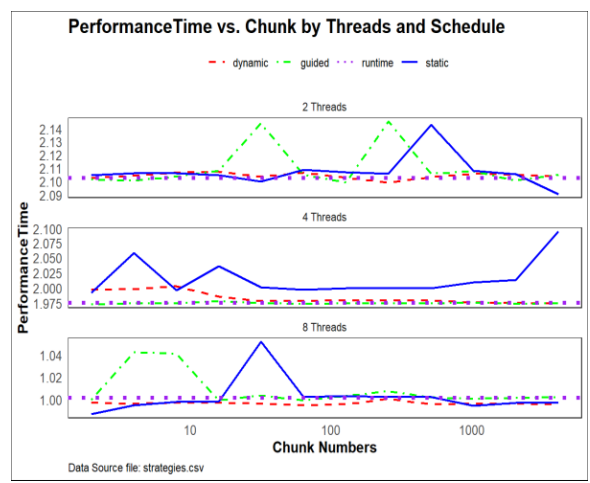

**Fig 4. PerformanceTime vs. Chunk by Threads and Schedule**

## 3.4 Experiment 3: Influence of Constructs on Program Runtime

This experiment primarily centered on three different constructs—Reduction Construct, Critical Partial Construct, and Critical Construct—and their influence on program runtime, given other variables remain constant. We compared these constructs under specific thread counts and scheduling strategies, conducting all tests under default parameter configurations. Here's a deeper analysis and conclusion based on our experimental data:

### 3.4.1 Analysis of Constructs' Impact on Runtime

Starting with Figure 5, we juxtaposed the runtime of

CITS5507 High Performance Computing

the "eat" module section using both the Reduction Construct and Critical Partial Construct. Evidently, the Reduction Construct excels in runtime for the eat module over the Critical Partial Construct. Since the eat module entails extensive computations and critical resource access, such a comparison becomes vital. Delving further, the Reduction Construct's runtime exhibited thread count independence, while the Critical Partial Construct's runtime showed an increase-then-decrease trend, consistent with findings from Experiment 1.

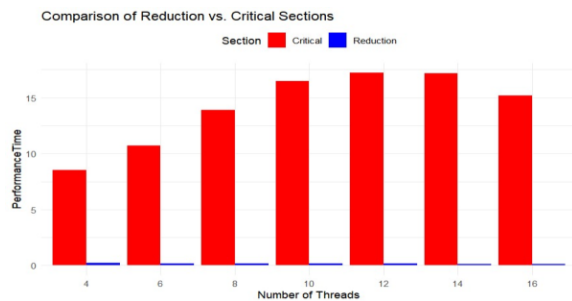

Fig 5. Comparison of Reduction vs. Critical Sections

Subsequently, as depicted in Figure 6, we compared the overall program runtime using the Reduction Construct, Critical Partial Construct, and Critical Construct. The outcomes underscored that the Reduction Construct consistently yielded optimal performance regardless of thread count settings, with efficiency amplifying the more it was utilized in the program.

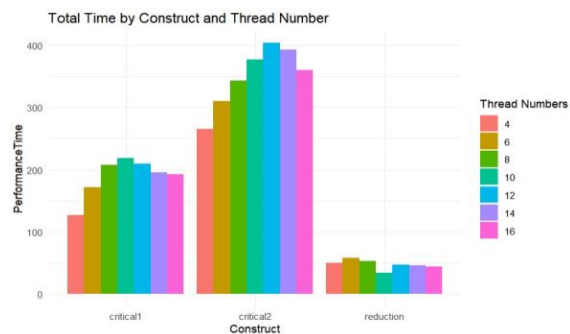

Fig 6. Comparison of Reduction vs. Critical Sections

## 4 Conclusion

Our study primarily revolves around the high-performance parallel optimization of the Fish School Behavior algorithm on the Setonix supercomputing platform using the OpenMP parallel computing framework. A systematic and in-depth analysis was conducted, especially for intricate, recurrent, and large-scale computational tasks. Delving into various factors of multithreaded parallelization such as thread counts, scheduling strategies, and OpenMP constructs, we drew from experimental outcomes to propose reliable parallel strategies and technical recommendations.

### 4.1 Key Findings

#### 4.1.1 Impact of Thread Count

Experiment one clearly demonstrated that the thread count significantly impacts the program's runtime. Notably, when employing the Reduction Construct, its performance surpasses that of the Critical Construct in most thread configurations, with a pronounced disparity at medium thread counts, such as 64 or 128.

#### 4.1.2 Choice of Scheduling Strategy

The second experiment emphasized the crucial role scheduling strategies play in determining performance. Specifically, when the thread count exceeds 20, the dynamic and guided scheduling strategies outperform the static one, showing a performance enhancement of up to 50%. Regarding specific chunk sizes, the dynamic scheduling strategy demonstrated commendable stability in most scenarios.

#### 4.1.3 Construct Selection

The third experiment reiterated the superior efficiency of the Reduction Construct. Be it within specific modules or the overall program execution, the Reduction Construct emerged as the top pick. Especially in the 'eat' module, which entails large-scale calculations and critical resource access, its dominance was evident.

**4.2 Practical Application Value**

Our findings not only offer insightful solutions and a deeper understanding of the Fish School Behaviour algorithm's parallel optimization on the Setonix platform but also serve as a valuable reference for other high-performance parallel computation optimization studies on the Setonix platform using the OpenMP framework. Particularly in scenarios involving repeated, large-scale computations, our study's methodology and conclusions bear significant relevance.

# 5 Future Prospects

Moving forward, we aim to delve deeper into other factors potentially affecting algorithm performance. Especially at the micro and macro levels, elements such as cache behavior and thread scheduling strategies within the operating system kernel, often overlooked in high-performance computing, hold pivotal importance for performance optimization. Additionally, given the diversity and complexity of algorithms, we plan to explore other parallel frameworks and techniques, aiming to provide more comprehensive and detailed optimization strategies for a myriad of computational needs.

Furthermore, we strongly believe our findings will continue to propel the research and application prospects in parallel computing. The unmatched advantages of high-performance parallel computing in tackling large-scale, complex real-world problems underpin our optimism. We anticipate future research to bring forth more innovations in this domain, ultimately offering more potent and efficient computational support to address practical challenges.

# References

[1] Sterling, Thomas, Maciej Brodowicz, and Matthew Anderson. *High performance computing: modern systems and practices*. Morgan Kaufmann, 2017.

[2] Vetter, Jeffrey S. "Contemporary high performance computing." *Contemporary High Performance Computing*. Chapman and Hall/CRC, 2017. 3-11.

[3] Hockney, Roger W., and Chris R. Jesshope. *Parallel Computers 2: architecture, programming and algorithms*. CRC Press, 2019.

[4] Marini, Federico, and Beata Walczak. "Particle swarm optimization (PSO). A tutorial." *Chemometrics and Intelligent Laboratory Systems* 149 (2015): 153-165.

[5] Diaz, Jose Monsalve, et al. "Evaluating support for openmp offload features." *Workshop Proceedings of the 47th International Conference on Parallel Processing*. 2018.

[6] Martineau, Matt, Simon McIntosh-Smith, and Wayne Gaudin. "Evaluating OpenMP 4.0's effectiveness as a heterogeneous parallel programming model." *2016 IEEE International Parallel and Distributed Processing Symposium Workshops (IPDPSW)*. IEEE, 2016.

[7] de Supinski, Bronis R., et al. "The ongoing evolution of openmp." *Proceedings of the IEEE* 106.11 (2018): 2004-2019.